\documentclass[aps,prd,twocolumn]{revtex4-2}

\usepackage[utf8]{inputenc} 
\usepackage[colorlinks]{hyperref}
\usepackage{amssymb,amsmath,mathtools,bm}
\usepackage{color}
\usepackage{graphicx}

\begin{document}

\title{Possibility of antiquark nuggets detection using meteor searching radars}

\author{V. V. Flambaum}
\affiliation{School of Physics, University of New South Wales, Sydney 2052, Australia}

\author{I. B. Samsonov}
\affiliation{School of Physics, University of New South Wales, Sydney 2052, Australia}

\author{G. K. Vong}
\affiliation{School of Physics, University of New South Wales, Sydney 2052, Australia}

\begin{abstract}
Within the quark nugget model, dark matter particles may be represented by compact composite objects composed of a large number of quarks or antiquarks. Due to strong interaction with visible matter, antiquark nuggets should manifest themselves in the form of rare atmospheric events on the Earth. They may produce ionized trails in the atmosphere similar to the meteor trails. There are, however, several features which should allow one to distinguish antiquark nugget trails from meteor ones. We study the properties of ionized trails from antiquark nuggets in the air and show that they may be registered by standard meteor radar detectors. Non-observation of such trails pushes up the mean baryon charge number in the quark nugget model, $|B|>4\times 10^{27}$.
\end{abstract}

\maketitle

\section{Introduction}

In Refs.~\cite{strangelet,nuclearite,Witten84} it was conjectured that SM quarks can combine and form stable compact composite objects characterized by a large baryon charge number $B\gg1$. Such objects are usually refereed to as quark nuggets (QNs) or strangelets, as their stability is stipulated by strange quarks. If stable quark nuggets exist, they would be perfect candidates for dark matter particles, because they are characterized by a very small cross section to mass ratio, $\sigma/M\ll 1\text{ cm}^2/\text{g}$, which ensures that they remain cosmologically `dark' despite strong interaction with visible matter.

A series of papers by A. Zhitnitsky and collaborators \cite{DW1,DW2,DW3,Zhitnitsky_2003} made significant progress in the quark nugget model of dark matter, wherein an axion-pion domain wall was proposed as a mechanism for stabilizing such objects. Another important feature introduced by A.~Zhitnitsky \cite{Zhitnitsky_2003,CCO} was separation of all dark matter particles into two families depending on whether they consist of matter (quarks) or antimatter (antiquarks), respectively. The idea behind this extension of the quark nugget model was that all antimatter in the Universe may be hidden inside the antiquark nuggets (anti-QNs), while the matter is represented by both QNs and visible baryonic matter. Thus, in total, the Universe has vanishing baryon number at all times. The difference between the numbers of QNs and anti-QNs  may be associated with a non-zero initial value of the QCD vacuum angle $\theta$ before it relaxes to zero value corresponding to the minimum of the potential energy.

The prediction of anti-QNs is a very attractive feature of the quark nugget model of dark matter, because these objects strongly interact with visible matter and may have interesting implications both in cosmology and in terrestrial observations. In Refs.~\cite{conjecture1,diffusexray,Lawson2007,WMAPhaze,Electrosphere,LAWSON201317,LAWSON2016,Lawson2017,FRB,QNinGalaxy1,QNradiation,thermal2021,FS3} it was argued that anti-QNs may be responsible for certain types of radiation observed in our Galaxy, while different possible atmospheric, seismic and cosmic ray-like events caused by anti-QNs on the Earth were discussed in Refs.~\cite{Budker2020,acoustic,Terrestrial4,Terrestrial3,Terrestrial2,Terrestrial1}. Importantly, the quark nugget model of dark matter is not just compatible with all these phenomena but explains them in a natural way within the frames of Standard Model of elementary particles. 

In this paper, we focus on observational properties of anti-QNs and study possibilities of their detection on the Earth.

One of the main parameters in the quark nugget model is the baryon charge number $B$, which quantifies mean number of quarks which comprise the nuggets. For convenience, in this paper we ignore sign of $B$, i.e.\ $B$ stands of the absolute value $|B|$ of the actual baryon charge. Other macroscopic parameters may be roughly expressed through $B$: the mean quark nugget radius is $R_0\approx 1\text{ fm}\times B^{1/3}$, and mean mass is $M\approx B m_p$, where $m_p$ is the proton mass. Note that the mean baryon number in this model is large, $B\gtrsim 10^{25}$, as lower values are excluded by non-observation of compact composite objects on such detectors as IceCube \cite{IceCube,BLimit} (see also Ref.~\cite{AQNReview} for a brief review). The corresponding limits on the size and mass are $R_0\gtrsim 2\times 10^{-5}$~cm and $M\gtrsim 15$~g. Even if such objects saturate local dark matter density $\rho_\text{DM}\sim 0.4\text{ GeV}/\text{cm}^3$, their number density $\rho_\text{DM}/M$  is very small, and probability of anti-QN passing through a detector is negligible. Therefore, we should look for possibilities of remote detection.

\begin{figure}[bth]
    \centering
    \includegraphics[width=0.48\textwidth]{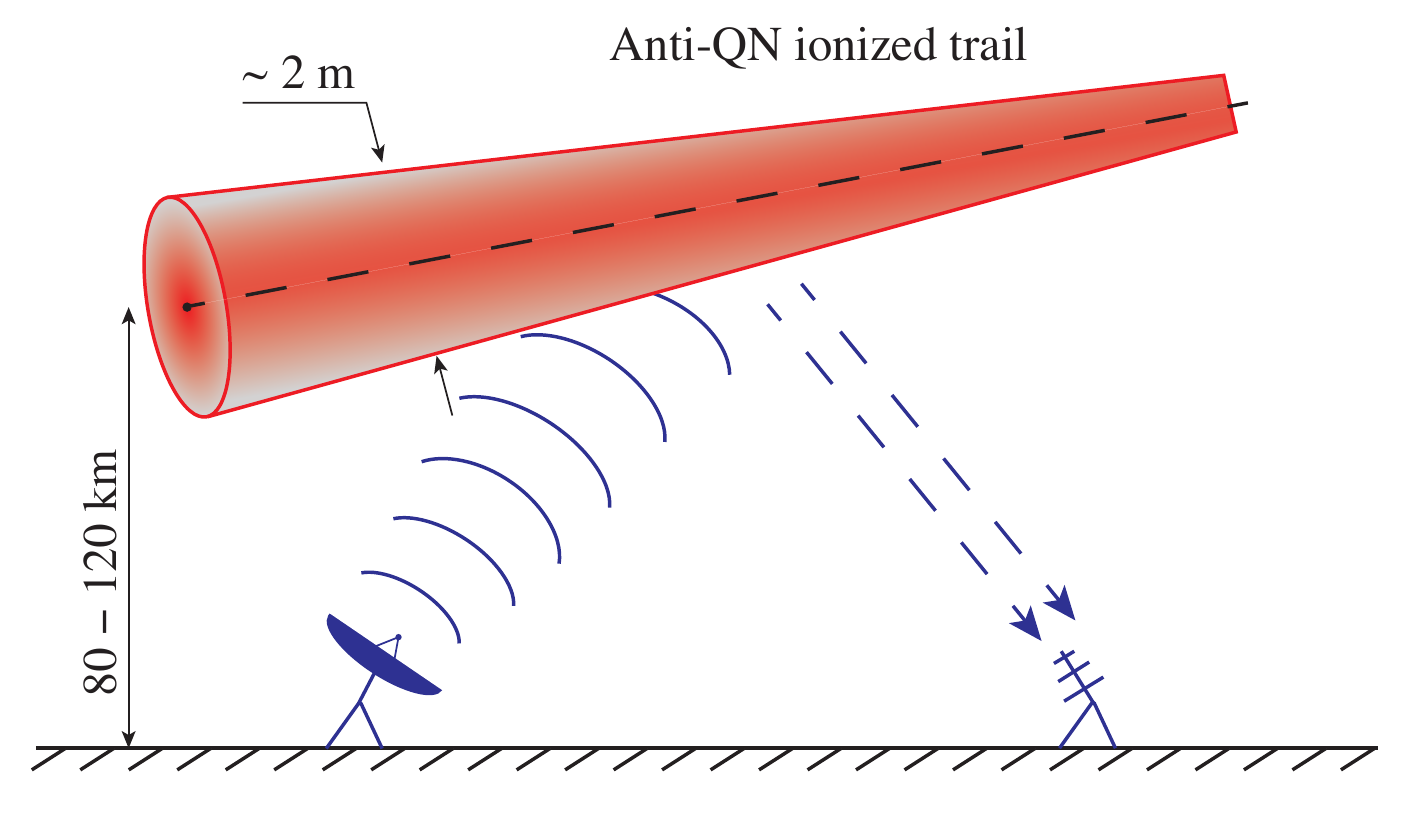}
    \caption{An illustration of radar detection technique of ionized anti-QN trails in the Earth's atmosphere (not to scale). Standard meteor radars may be suitable for detection of anti-QN trails.}
    \label{fig:Radar}
\end{figure}

In Refs.~\cite{QNradiation,thermal2021} it was shown that an anti-QN moving through a medium emits specific types of radiation which includes x-rays, $\gamma$-rays, and fast ionizing particles such as $\pi^\pm$ pions, (anti)muons, electrons and positrons. In this paper, we study ionizing properties of this radiation from anti-QNs in the Earth's atmosphere. We estimate the density of electrons in the atmosphere produced by this radiation and compare it with the one from ordinary meteors. We show that at altitude $h\sim 80-120$ km above sea level the anti-QNs produce an ionized trail which is comparable to the one from typical meteors. Thus we conclude that meteor radar detection technique may be suitable for detection of anti-QNs. This technique is scematically shown in Fig.~\ref{fig:Radar}.

A long-standing problem in cosmology is direct detection of dark matter particles in laboratory. In the case of quark nugget model, the Earth itself may be considered as a detector. This idea was recently advertised in Ref.~\cite{DMradar} where it was proposed to search for compact composite objects in the Earth's atmosphere using meteor radar detectors. Non-observation of such objects allowed the authors of this work to place limits on the mass and cross section of dark matter particles. The results of the work \cite{DMradar} do not, however, apply to the quark nugget dark matter because anti-QNs have a different mechanism of air ionization. In this paper, we fill this gap and study the limits on the mean baryon charge number in the quark nugget model from existing meteor radar observations.

The rest of this paper is organised as follows. In the next section, we briefly overview the properties of thermal and annihilation radiations in the quark nugget model studied earlier in Refs.~\cite{QNradiation,thermal2021}.
In Sec. \ref{ionisation}, we consider ionizing properties of anti-QNs travelling through the Earth's atmosphere and calculate the electron density produced by different types of radiation from an anti-QN. In Sec. \ref{detection}, we estimate the width of ionized trail generated by an anti-QN in the air and compare it with the one from conventional meteors. In Sec.~\ref{SecOther}, we discuss specific properties of anti-QN trails which may help distinguish them from the meteor ones and detect them in the future.
Sec.~\ref{summary} is devoted to a summary and discussion of the obtained results. In Appendices, we present some technical details of our calculations and tables with numerical results employed in our estimates.

Throughout this paper we use natural units with $\hbar=1$, $c =1$ and $k_{B}=1$.

%%%%%%%%%%%%%%%%%%%%%%%%%%%%%%%%%%%%%%%%%%%%%%%%%

\section{Thermal and annihilation radiation from antiquark nuggets passing through the atmosphere}
\label{thermal}

Anti-QNs interacting with nuclei in the Earth's atmosphere could result in the emission of observable thermal radiation. In this section, we briefly overview on the properties of thermal radiation from anti-QNs as studied in Ref.~\cite{thermal2021}. Then, we study properties of the thermal radiation from anti-QNs passing through the atmosphere.

\subsection{Thermal emissivity of antiquark nuggets}
The main assumption of the QN model is that quark nuggets are composed of a large number of quarks or antiquarks. It is usually assumed that the density of the quark nugget core is slightly higher than the nuclear matter density. This assumption is quantified by the following relation between the anti-QN radius $R_0$ and the baryon charge number $B$
\begin{equation}
        R_{0} \simeq B^{1 /3} \times 1 \mbox{ fm}\,. \label{radius0}
\end{equation}

The baryon charge number should be sufficiently large to ensure that anti-QNs survive till the present days and be still potentially observable \cite{sizedistribution,FS3} (see also \cite{AQNReview} for a review)
\begin{equation}
      B\gtrsim 10^{25}\,.      
      \label{Bold}
\end{equation}
In particular, for $B=10^{25}$, Eq.~(\ref{radius0}) implies $R_{0} \simeq 2.2\times 10^{-7} \mbox{ m}$.

The antiquark core possesses a large electric charge which is neutralized by the charge of the positron cloud around the antiquark core. At zero temperature, these positrons are in a degenerate state with Fermi energy in the range 10-100 MeV \cite{strangestars}. At non-zero temperature, the positrons are excited above the Fermi surface and obey the Fermi-Dirac distribution. The density distribution in the positron cloud was studied in Ref.~\cite{QNradiation}, and the model of thermal radiation from the positron cloud was developed in Ref.~\cite{thermal2021}. In this model, thermal radiation is produced by fluctuations of density in the positron cloud, similar to plasma oscillations producing thermal radiation in small metallic particles. However, in contrast with the latter, because of a very high density in the positron cloud, anti-QNs possess relatively high plasma frequency, $\omega_p\simeq 2$~MeV \cite{thermal2021}. Given this plasma frequency, we introduce a (complex) dielectric constant 
\begin{equation}
        \varepsilon(\omega) = 1- \frac{ {\omega_{p}}^{2}}{\omega^2+i\gamma\omega}\,,
        \label{epsilon}
\end{equation}
where $\gamma$ is a damping constant. In Ref.~\cite{thermal2021}, the damping constant was estimated with the use of Drude model, $\gamma \simeq 0.5 \mbox{ keV}$.

In general, the radiation power per unit surface area of anti-QN per unit frequency interval may be written as
\begin{equation}
        P(\omega,T) = \pi E(\omega) I_{0}(\omega,T)\,,
        \label{Pbb}
\end{equation}
where 
\begin{equation}
        I_{0}(\omega,T) = \frac{\hbar \omega^{3}}{4\pi^{3} c^{2}} \frac{1}{\exp\left(\hbar\omega /(k_{B}T)\right)-1}
\end{equation}
is the Planck function and $E(\omega)$ is a thermal emissivity coefficient. The latter may be calculated within the Mie theory which describes the scattering and absorption of light on compact bodies, see, e.g., \cite{mietheory}. For a spherical particle, the thermal emissivity function is given by a series expansion over  spherical harmonics. For high frequencies,  $c/R_0 \ll \omega \ll \omega_p$, the expression for the thermal emissivity coefficient simplifies drastically, 
\begin{equation}
        E(\omega) \approx 5.36\operatorname{Re}\left[ \varepsilon(\omega) \right]^{-1 /2}\, ,
        \label{Ehigh}
\end{equation}
where $\varepsilon(\omega)$ is the complex dielectric constant (\ref{epsilon}), and the coefficient 5.36 was calculated numerically in Ref.~\cite{thermal2021}. Remarkably, the function (\ref{Ehigh}) as well as the radiation power  per unit surface area in Eq.~(\ref{Pbb}) are independent of the anti-QN radius $R_0$ and, thus, of the baryon charge number $B$. This feature is a result of the short wavelength approximation $\lambda \ll R_0$ and it is somewhat similar to that for the black body radiation (BBR) where the radiation is defined by the temperature $T$ rather than specific content of the body. In contrast with BBR, the anti-QN radiation depends also on the plasma frequency $\omega_p$ and damping constant $\gamma$.

The total radiation power per unit surface area of the quark nugget can be found by integrating Eq.~\eqref{Pbb} over $\omega$,
\begin{equation}
        F_\text{rad}(T) = \int_{0}^{\infty} \pi E(\omega) I_{0}(\omega,T) \, d\omega.
        \label{Frad}
\end{equation}
The thermal radiation spectrum of one quark nugget is then found by multiplying Eq.~\eqref{Frad} by the quark nugget surface area $4\pi {R_{0}}^2$.

\subsection{Parameters of thermal radiation from anti-QNs in the air}
\label{SecXrayEmission}

When an anti-QN passes through the air, it annihilates  air molecules releasing roughly $2m_p c^2$ of energy per each annihilated nucleon, where $m_p\approx 938\text{ MeV}/c^2$ is the proton mass. Some fraction $\eta$ of this released energy will heat the positron cloud to a temperature $T$, resulting in the emission of thermal radiation with power (\ref{Pbb}). It is hard to accurately estimate the value of the parameter $\eta$ because the model of matter annihilation on anti-QNs is not sufficiently developed. In Ref.~\cite{QNradiation} we assumed that this annihilation process is similar to the proton-antiproton annihilation with emission of charged and neutral $\pi$ mesons. However, it is unclear whether this annihilation happens just on the boundary of anti-QN or deep inside it. In the former case, approximately 50\% of produced pions are emitted outside the anti-QN  while the other 50\% thermalize inside the anti-QN core. In this case, $\eta\approx 0.5$. If the annihilation happens deep inside the anti-QN core, most of the produced pions are absorbed and the released energy thermalizes. This case corresponds to $\eta\approx 1$. 

In this paper, we will consider two limiting cases with $\eta=0.5$ and $\eta=1$, respectively, meaning that the actual value of this parameter is within this range. As we will show, all results and conclusions of this paper weakly depend on the actual value of this parameter. 

When an anti-QN moves through the air with density $\rho_{\text{air}}$, it acquires effective internal temperature $T$ due to the matter-antimatter annihilation. As a result, it radiates with the power 
\begin{equation}
W_\text{rad}=4\pi R_0^2 F_{\text{rad}}(T)\,,
\label{Wrad}
\end{equation}
where $F_{\text{rad}}$ is given by Eq.~(\ref{Frad}). The incoming energy flux due to the air molecule annihilation on the anti-QN is 
\begin{equation}
W_\text{in} = 2m_p\eta\sigma_{\text{ann}}n_{\text{air}}v\,,
\label{Win}
\end{equation}
where $\eta$ is the fraction of thermalized energy released in the nucleon annihilation, $\sigma_\text{ann}$
is the annihilation cross section for air molecules on anti-QN \cite{QNradiation}, $n_{\text{air}} = \rho_{\text{air}}/m_p$ is the nucleon number density in the air and $v\approx 10^{-3}c$ is the anti-QN velocity in the air. 

The annihilation cross section may written as
\begin{equation}
    \sigma_{\text{ann}}= \kappa \pi R_0^2\,,
    \label{sigma}
\end{equation}
where $\kappa$ is a suppression coefficient which is not known exactly. We can roughly estimate this coefficient as follows. Fist note that for a single proton scattering off anti-QN $\kappa\approx 1$, because the proton is trapped near the anti-QN core by the Coulomb attraction and  loss of the proton kinetic energy due to friction within the positron cloud \cite{QNradiation}. For larger atoms like nitrogen or oxygen this coefficient is unlikely to exceed the value of 0.5, because neutrons from the nucleus are not confined with the antiquark core by the Coulomb attraction. Note that the anti-QN Coulomb field accelerates charged nuclei leading to the potential energy inside the quark core  of order 30 MeV per proton that exceeds nucleon binding energy in an incident nucleus. After a collision with anti-QN core the incident nitrogen or oxygen nucleus is likely to be broken into unbind nucleons near the antiquark core boundary. 

The suppression coefficient $\kappa$ may be less than 0.5 if the annihilation of the first few protons from incident nucleus transfers much kinetic energy to the rests of this nucleus sufficient for their escape without annihilation. Thus, it plausible that the suppression coefficient may be of order
\begin{equation}
    \kappa \approx 0.25\,,
    \label{kappa}
\end{equation}
but is unlikely to be much smaller than that. 

Although the above arguments on the value of the coefficient $\kappa$ are just qualitative, they allow us to perform further estimates within an order of magnitude accuracy. In fact, it is possible to show that the final conclusions of this work weakly depend on the particular value of the suppression parameter $\kappa$, and we will assume the value (\ref{kappa}) in what follows. 

Equating the incoming (\ref{Win}) and radiated (\ref{Wrad}) energy fluxes through the anti-QN surface, we find the equation which determines the effective anti-QN temperature in the air,
\begin{equation}
         F_\text{rad}(T)  = \frac{\eta\kappa}2  \rho_\text{air}  v\,.
        \label{Teq}
\end{equation}

\begin{figure}
    \centering
    \includegraphics[width=8.5cm]{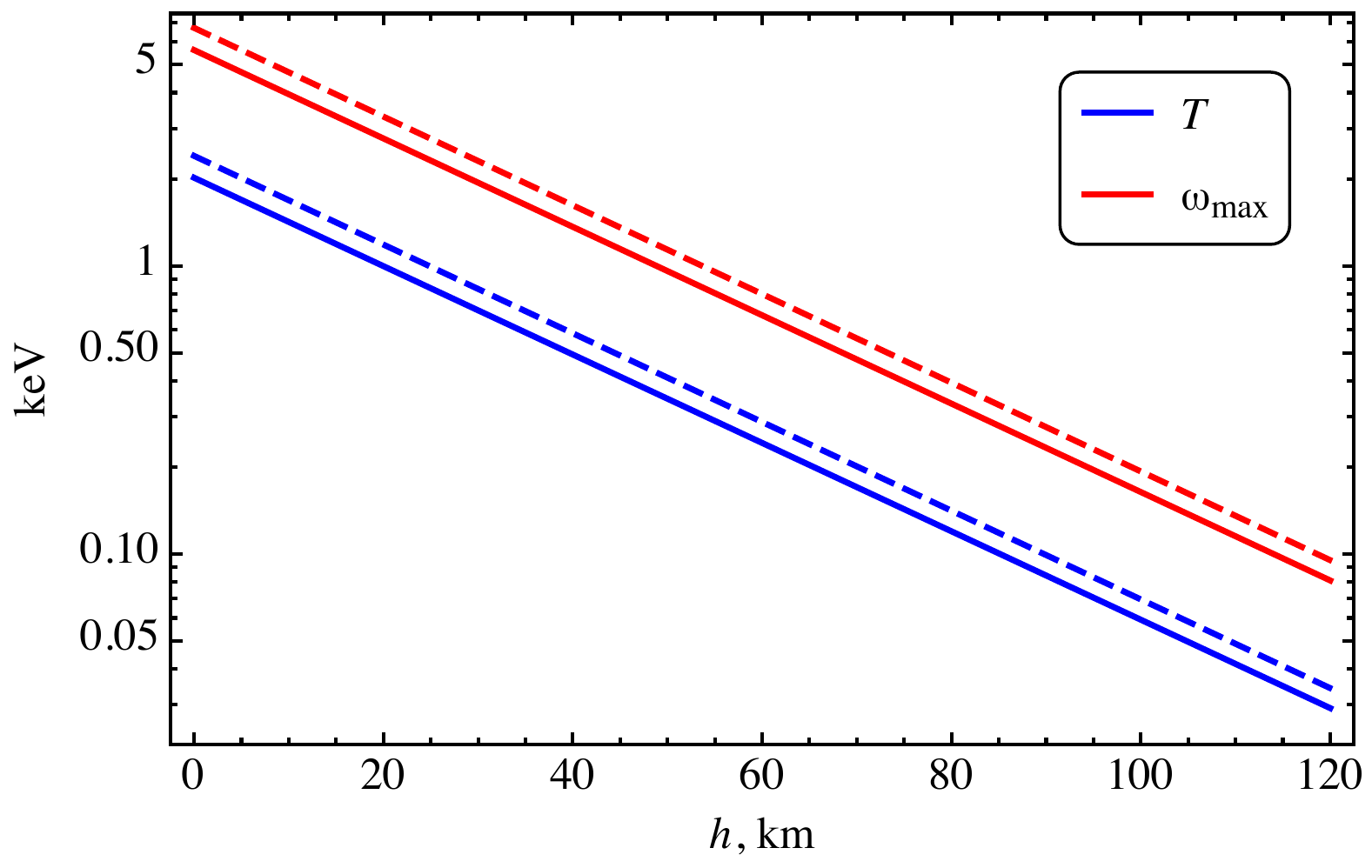}
    \caption{Plots of the effective anti-QN temperature $T$ (blue lines) and radiation peak angular frequency $\omega_\text{max}$ (red lines) as functions of the altitude $h$ above the sea level. Solid and dashed lines correspond to $\eta=0.5$ and $\eta=1$, respectively, where $\eta$ is a fraction of the thermalized energy in the process of annihilation of air molecules on anti-QNs.}
    \label{FigT}
\end{figure}

Within the isothermal atmospheric model, the air density as a function of the altitude above the sea level $h$ may be approximately described by the following exponential function
\begin{equation}
    \rho_\text{air}(h) = \rho_0 e^{-h/h_0}\,,
    \label{AirDensity}
\end{equation}
where $\rho_0 = 1.2\text{ kg/m}^3$ is the air density at the sea level and $h_0 = 7\text{ km}$. With the air density given by the function (\ref{AirDensity}) and the thermal radiation power represented by Eq.~(\ref{Frad}), equation (\ref{Teq}) may be solved numerically for any particular altitude. As a result, we find the effective anti-QN temperature as a function of the altitude $h$,
\begin{equation}
    T(h) = T_0 e^{-h/h_1}\,,\quad
    T_0=\left\{
    \begin{array}{l}
        7.55\mbox{ keV for }\eta = 0.5 \\
        9.00\mbox{ keV for }\eta = 1\,,
    \end{array}
    \right.
    \label{Teff}
\end{equation}
where $h_1 = 28.2$~km. Numerical values of the effective temperature (\ref{Teff}) are presented in Table~\ref{TableT} for different altitudes, and the corresponding plot is given in Fig.~\ref{FigT}.

\begin{table}[tb]
    \centering
    \begin{tabular}{c|c|c|c|c|c|c}
& \multicolumn{3}{c|}{$\eta=0.5$} &
\multicolumn{3}{c}{$\eta=1$}
    \\\hline
  $h/$km & $T/$keV & $\omega_{\text{max}}/$keV & $\bar\omega/$keV &$T/$keV & $\omega_{\text{max}}/$keV & $\bar\omega/$keV \\\hline\hline
 0  & 7.55 & 21.0 & 29.2 & 9.00 & 25.0 & 34.7\\
 20 & 3.73 & 10.3 & 14.3 & 4.43 & 12.3 & 17.0\\
 40 & 1.84 & 5.10 & 7.00 & 2.18 & 6.06 & 8.32\\
 60 & 0.906 & 2.52 & 3.44 & 1.07 & 2.98 & 4.09\\
 80 & 0.447 & 1.24& 1.70 & 0.527& 1.47 & 2.02\\
 100 & 0.221 & 0.612& 0.857 &0.259&0.722 & 1.01\\
 120 & 0.109 & 0.302& 0.441 &0.128&0.355 &0.517\\\hline\hline
    \end{tabular}
    \caption{Values of the effective anti-QN temperature $T$, radiation peak and mean angular frequencies $\omega_\text{max}$ and $\bar\omega$, respectively, as functions of the altitude $h$ above the sea level. These functions are given for two values of the parameter $\eta$ which quantifies the fraction of thermalized energy in the full energy released upon the nucleon annihilation on anti-QN.
    }
    \label{TableT}
\end{table}

The effective anti-QN temperature $T$ specifies the radiation spectrum from anti-QNs in the atmosphere (\ref{Pbb}). This spectrum differs from the black body one by the emissivity coefficient $E(\omega)$. Let $\omega_\text{max}$ be a peak frequency corresponding to the maximum of the function (\ref{Pbb}) for each value of the effective temperature $T$. Consider also the mean angular frequency defined as
\begin{equation}
    \bar\omega = \frac{\int_0^\infty \omega P(\omega,T)d\omega}{\int_0^\infty  P(\omega,T)d\omega}\,,
\end{equation}
where $P(\omega,T)$ is given by Eq.~(\ref{Pbb}). Since the effective anti-QN temperature is the function of the altitude (\ref{Teff}), both $\omega_\text{max}$ and $\bar\omega$ depend on the altitude, $\omega_\text{max}=\omega_\text{max}(h)$ and $\bar\omega=\bar\omega(h)$. These functions may be calculated numerically using Eqs.~(\ref{epsilon},\ref{Pbb},\ref{Ehigh}) and (\ref{Teff}). Their values are represented in Table~\ref{TableT} and are plotted in Fig.~\ref{FigT}. We find also that the dependence of the peak and average frequencies on the altitude may be approximated by the following exponential functions:
\begin{align}
    \omega_\text{max}(h) &= \omega_0 e^{-h/h_1}\,,\quad
    \omega_0=\left\{
    \begin{array}{l}
        21.0\mbox{ keV for }\eta = 0.5 \\
        25.0\mbox{ keV for }\eta = 1\,,
    \end{array}
    \right. 
    \\
    \bar\omega(h) &= \bar\omega_0 e^{-h/h_1}\,,\quad
    \bar\omega_0=\left\{
    \begin{array}{l}
        29.2\mbox{ keV for }\eta = 0.5 \\
        34.7\mbox{ keV for }\eta = 1\,,
    \end{array}
    \right.
\end{align}
where $h_1=28.2$ km. Note that the thermal radiation from anti-QN in the Earth's atmosphere falls within the x-ray spectrum. As we will show below, this radiation represents the main source of ionizing radiation from anti-QNs.

We have considered the functions $\omega_\text{max}(h)$, $\bar\omega(h)$ and $T(h)$ for the two limiting values of the parameter $\eta$ which specifies the fraction of the thermalized energy in the process of annihilation of air molecules on anti-QNs. As is seen from Table~\ref{TableT}, the variations of the values of $\omega_\text{max}$, $\bar\omega$ and $T$ do not exceed 20\% between the values $\eta=0.5$ and $\eta=1$. Thus, the pattern of thermal radiation from anti-QNs weakly depends on $\eta$. 

\subsection{Direct annihilation radiation from anti-QNs}
\label{SecAnnihilation}

The annihilation radiation from anti-QNs was studied in Ref.~\cite{QNradiation}. In this subsection, we briefly overview basic properties of this radiation.

As is demonstrated in Ref.~\cite{QNradiation}, anti-QNs possess a strong electric field near the boundary which is able to ionize neutral atoms and molecules colliding with anti-QNs. The ionized electrons are repelled off the anti-QN while the positively charged nuclei are attracted to the antiquark core. In the case of hydrogen, proton loses energy due to friction in the positron gas  and is trapped inside the core. Thus, proton  inevitably annihilates inside the quark core releasing about 2 GeV of energy. A possibility of transformation of proton to neutron, which can escape, does  not affect the result significantly. As it was explained above, heavier nuclei may decay to protons and neutrons in the process of collisions with the core, and the annihilation may be incomplete reducing released energy by an order of magnitude.  Note also that atomic electrons do not play much role in this process and may be ignored. 

Exact pattern of annihilation of atomic nuclei inside antiquark matter is not known because the state of the latter is not well studied. To make our estimates, here we assume that the pattern of this annihilation is similar to the proton-antiproton annihilation, although this assumption may be revisited in future works. This assumption means that primary annihilation products are charged and neutral $\pi$ mesons. Although there are many decay channels in the proton-antiproton annihilation \cite{ppbratios}, on average five pions are produced among which three are charged $\pi^\pm$ mesons and two are neutral $\pi^0$. Each pion has total energy (including its rest mass) of about 375 MeV which comes out as one fifth of the total $2m_pc^2$ energy. Thus, these pions are weakly relativistic.

In Ref.~\cite{QNradiation} we assumed that protons  annihilate very close to the boundary of the antiquark core, and only about 50\% of produced pions are emitted outside the antiquark core and escape while the other 50\% of pions go inside the antiquark core where they thermalize and further get absorbed via strong interactions. This corresponds to $\eta=0.5$, where $\eta$ is the fraction of thermalized pions. However, it is not excluded that the nucleus penetrates deep inside the antiquark core before direct annihilation happens. In this case $\eta$ is close to 1, and the fraction of emitted outside pions is $(1-\eta)$. In this case, the annihilation radiation from anti-QNs is suppressed. For the sake of generality, the annihilation radiation will be accounted for with the factor $(1-\eta) \kappa$,  where $\kappa$ is the annihilation suppression coefficient (\ref{kappa}).

Let us consider pions produced in the process of annihilation of nuclei on anti-QNs. Neutral pions are very short-living, with main decay channel into two $\gamma$ photons, $\pi^0\to 2\gamma$. The energies of this photons are close to 200 MeV, with some distribution around this value, depending on the energy of original $\pi^0$. Thus, we can roughly estimate that each nucleon annihilating on anti-QN produces $4(1-\eta)\kappa$ gamma photons with typical energy around 200 MeV.

Charged $\pi^\pm$ mesons have lifetime $\tau = 2.6\times 10^{-8}$~s and decay into (anti)muons and muonic (anti)neutrinos, $\pi^\pm \to \mu^\pm + \nu_\mu(\bar\nu_\mu)$. Considering that some fraction of the pion energy is taken by (anti)neutrino, the (anti)muon energy (including its rest mass) should be around 295 MeV. Muons further decay into electrons and muonic neutrino plus electronic antineutrino, $\mu^- \to e^- + \bar\nu_e + \nu_\mu$, $\mu^+ \to e^+ + \nu_e + \bar\nu_\mu$, with lifetime $\tau=2.2\,\mu$s. As a result, the ultrarelativistic electrons and positrons, as well as the corresponding (anti)neutrinos, are final decay products in the anti-QN annihilation process. Each nucleon annihilation thus produces about $3(1-\eta)\kappa$ (anti)muons which further decay into the same number of electrons (positrons). All these particles ionize air molecules in the Earth atmosphere.

%%%%%%%%%%%%%%%%%%%%%%%%%%%%%%%%%%%%%%%%%%%%%%%%%%%%%%%%%%%%%%%%%%%%%%%%%%%%%%%%%%%%%%%%

\section{Electron density in the antiquark nugget trail}\label{ionisation}

As an anti-QN passes through the air, air molecules annihilate in the antiquark core with emission of ionising particles considered in the previous section. These particles ionise the surrounding air, producing a trail of free electrons and ions along the anti-QN path. We call this the ``anti-QN trail" by analogy with ionised meteor trails. 

An important measurable quantity in meteor trails is the free electron density $n_e$. Therefore our goal in this section is to estimate the electron density in the anti-QN trail and compare it with the meteor one. In the next section, we will consider the possibility of detection of anti-QN trails with meteor radars. Note that it is sufficient to study only the density of free electrons since ions are much heavier and make a negligible contribution to radar observations in the atmosphere \cite{mckinley1961,scattering}.

It should be noted that trails from quark nuggets (but not anti-QNs) in the atmosphere were studied in Refs.~\cite{nuclearite,DMradar}. These trails are formed in the process of direct scattering with air molecules rather annihilation. We emphasise that the focus of the present paper is on the trails from \textit{antiquark} nuggets which have not been explored yet.

\subsection{Initial electron density distribution}

In general, an anti-QN enters the Earth's atmosphere at a zenith angle $\theta$ and produces an ionized trail which has a shape of a tapered cylinder. At any given altitude $h$, the cross section of this trail is approximately circular. At each altitude, we want to study the free electron density in the ionized trail, $n_e=n_e(a)$, as a function of distance $a$ from the anti-QN trajectory (axis of the cylinder). 

As is shown in the previous section, the annihilation of air molecules on the anti-QNs yields the emission of different ionizing particles, including x-rays, charged pions, muons, $\beta$ and $\gamma$ particles. Each of this type of particles gives a contribution $n_i(a)$ to the total electron density,
\begin{equation}
    n_e(a) = \sum_i n_i(a)\,.
    \label{neSum}
\end{equation}
The summation index $i$ labels different types of the ionizing particles. Note that each type of these particles is specified also by an attenuation length in the air $L_i$ and mean initial (kinetic) energy ${\cal E}_i$ which is spent for ionization of the air. These quantities depend on the altitude $h$ through the air density $\rho_\text{air}(h)$.

In Appendix \ref{AppA}, it is demonstrated that at any given altitude the contribution from $i$-th ionizing species to the electron density may be approximated by the function
\begin{equation}
    n_i(a) = \frac{q_i}{2\pi L_i a }\tan^{-1}\sqrt{ \frac{L_i^2}{a^2} - 1}\,,
    \label{ne-i}
\end{equation}
where $q_i\equiv 2\pi\int_0^{L_i} n_i(a)a\,da$ is the column density in the ionized trail. According to Eq.~(\ref{q}), the column density may be expressed via the production rate of $i$-th type of ionizing particles $W_i$, mean number of produced electrons per one ionizing particle $N_i$ and the anti-QN typical velocity $v\approx 10^{-3}c$:
\begin{equation}
    q_i = \frac{W_i N_i}{v}\,.
    \label{qi0}
\end{equation}

The number $N_i$ may be represented as 
\begin{equation}
    N_i = \frac{{\cal E}_i}{I}\,,
\end{equation}
where ${\cal E}_i$ is the mean (kinetic) energy of $i$-th ionizing particle and $I$ is the mean energy required to produce one electron in the air ionization process. In Ref.~\cite{Wair}, it is shown that $I\approx 33$ eV for a variety of fast ionizing particles, including $\alpha,\beta,\gamma$ particles and x-rays. Thus, the electron line density (\ref{qi0}) depends mainly on the ionizing particles mean kinetic energy ${\cal E}_i$ and their production rate $W_i$,
\begin{equation}
    q_i = \frac{W_i {\cal E}_i}{vI}\,.
    \label{q-i}
\end{equation}

Note that the quantities $L_i$, $W_i$ and ${\cal E}_i$ depend on the altitude via the air density $\rho_\text{air}(h)$. Thus, the electron volume and line densities (\ref{ne-i}) and (\ref{q-i}) are effectively functions of the altitude $h$ as well.

\subsection{Contributions to the electron density}

In this subsection, we determine the pairs $(q_i,L_i)$ for different types of ionizing radiation from anti-QNs. This will allow us to find the function $n_e(a)$ from Eqs.~(\ref{neSum}) and (\ref{ne-i}).

\subsubsection{X-ray emissions}
As was shown in Sec.~\ref{SecXrayEmission}, anti-QNs emit x-rays in continuum spectrum in the range roughly from 1 to 100 keV. This emission is sourced by thermal fluctuations in the positron cloud in anti-QN. For the thermal radiation, the numerator in Eq.~(\ref{q-i}) is proportional to the annihilation rate of air molecules with the coefficient $\eta\kappa$,
$W_\text{x} {\cal E}_\text{x} = 2m_p \eta \kappa \pi R_0^2 v n_\text{air}$. Considering that $m_p n_\text{air} = \rho_\text{air}$, we obtain the following relation for the electron line density contribution from x-rays:
\begin{equation}
    q_\text{x}(h) = \frac{2\eta \kappa\pi R_0^2 \rho_\text{air}(h)}{I}=(7.6\times 10^{20}\text{m}^{-1})e^{-h/h_0}\,,
    \label{q-x}
\end{equation}
where $h_0 = 7$~km. Here we assumed the value $\eta\kappa=0.1$ in accordance with Eq.~(\ref{kappa}).

The absorption of x-rays in the air is well studied, see, e.g., \cite{PDG}. Let $L_\text{x0}$ be attenuation length of x-rays with mean energy $\bar\omega$ in the air at the sea level. Then, the attenuation length at arbitrary altitude $h$ may be written as
\begin{equation}
    L_\text{x}(h) = L_\text{x0}\frac{\rho_\text{air}(0)}{\rho_\text{air}(h)} =L_\text{x0} e^{h/h_0}\,.
    \label{L-x}
\end{equation}

In particular, at the altitude $h=100$ km, the mean x-ray energy is $\bar\omega = 0.857$ keV, see Table~\ref{TableT}. The corresponding attenuation length and column density are $L_\text{x} = 2.4$ km and $q_\text{x} = 4.7\times 10^{14}$ electrons/m. For other altitudes, these values may be found with Eqs.~(\ref{q-x}) and (\ref{L-x}).

In the following subsection we will estimate contributions to the electron density from the products of direct annihilation. As will be shown below, these products have large  attenuation lengths and their contributions to the volume electron density are significantly smaller than the contribution of the thermal x-rays at distance $a<L_\text{x}$ from the anti-QN trajectory.

\subsubsection{$\gamma$ photons}
As per discussion in Sec.~\ref{SecAnnihilation}, anti-QNs emit $\gamma$ photons with mean energy ${\cal E}_\gamma = 200$ MeV and production rate $W_\gamma = 4(1-\eta)\sigma_\text{ann}n_\text{air} v$, where $\sigma_\text{ann}\approx \kappa\pi R_0^2$ and $n_\text{air} = \rho_\text{air}/m_p$. With this production rate, the electron line density (\ref{q-i}) reads
\begin{equation}
    q_\gamma(h) = \frac{4\kappa(1-\eta)\pi R_0^2 n_\text{air}{\cal E}_\gamma}{I}
    =(3.2\times 10^{20}\text{m}^{-1})e^{-h/h_0}\,.
    \label{q-gamma}
\end{equation}
In particular, at the altitude $h=100$ km, the contribution to the line density from $\gamma$ photons is about two times smaller than that from x-rays: $q_\gamma =2\times 10^{14}$ electrons/m. 

The attenuation length for $\gamma$ photons is
\begin{equation}
    L_\gamma(h) = L_{\gamma0}\frac{\rho_\text{air}(0)}{\rho_\text{air}(h)}= L_{\gamma0} e^{h/h_0}\,,
    \label{L-gamma}
\end{equation}
where $L_{\gamma0}\approx 0.5$ km is the attenuation length of $\gamma$ photons with energy ${\cal E}_\gamma = 200$ MeV at sea level. 

Note that at high altitudes the above formula (\ref{L-gamma}) gives unreasonably high attenuation length of $\gamma$ photons, because it does not take into account variations of the air density along the photon path. In particular, at $h=100$ km, $L_\gamma \sim 750000$ km. This length is much higher than the considered altitude, and, this result should be taken with care. It just shows that column density $q_\gamma$ is spread over a large range around the anti-QN trajectory, and the corresponding volume density contribution $n_\gamma$ is much smaller than that from x-rays considered above.

\subsubsection{Muons and $\pi^\pm$ mesons}

As is shown in Sec.~\ref{SecAnnihilation}, anti-QNs emit charged pions as a result of air molecules annihilation in the antiquark core. The mean kinetic energy of these pions was estimated as $E=235$ MeV, so these are weakly relativistic particles with $\beta \sim 0.9$. Given the lifetime of $\pi^\pm$ is $\tau = 2.6\times 10^{-8}$~s, their decay length is about $l_\text{decay}=20$ m, behind which they are likely to decay into (anti)muons and (anti)neutrino.

The mean kinetic energy of the produced (anti)muons is $E = 190$ MeV, so these particles are also weakly relativistic with $\beta\sim 0.9$. Thus, the ionizing properties in the air of pions and muons are similar, and they may be considered in a unified manner. The decay length (in vacuum) of a muon with such kinetic energy is about $l_\text{decay}=1.7$ km that is much larger than the pion mean free path. Therefore, the latter may be neglected.

When a muon moves through the air along the direction $x$, it loses its kinetic energy $E$ with the rate $dE/dx$, and its stopping range may be denoted as $l_\text{stop}$. These quantities are well studied, see, e.g., Ref.~\cite{MuonStoppingPower}. We present the values of these functions in Appendix \ref{AppMuon} in the range of energies from 1 to 300 MeV. With these data it is possible to show that at altitudes $h\gtrsim 10$ km the muon stopping range is greater than its decay length, $l_\text{stop}>l_\text{decay}$. Thus, at these altitudes muons decay before they are stopped in the atmosphere. Therefore, the ionization range in the air due to the muons is
\begin{equation}
L_\mu \equiv l_\text{decay} = 1.7\text{ km}\,.
\label{L-mu}
\end{equation}

The energy lost by a muon along the path $L_\mu$ in the air is
\begin{equation}
    {\cal E}_\mu = \int_0^{L_\mu} (-d E/dx) dx\,.
\end{equation}
Note that this energy is lost predominantly to the ionization of air molecules \cite{MuonStoppingPower}. Therefore, making use of the muon production rate in the annihilation of air molecules on anti-QN, $W_\mu = 3\kappa(1-\eta)\pi R_0^2 n_\text{air} v$, we find the corresponding electron line density (\ref{q-i}):
\begin{equation}
    q_\mu = 3\kappa(1-\eta) \pi R_0^2 n_\text{air} \frac{{\cal E}_\mu}{I}\,.
    \label{q-mu0}
\end{equation}
Using the numerical values of $d E/dx$ from Appendix~\ref{AppMuon}, we find that the electron line density (\ref{q-mu0}) is well described by the following function:
\begin{equation}
    q_\mu(h) = (4.7\times 10^{20}\text{m}^{-1}) e^{-2h/h_0}\,,
    \label{q-mu}
\end{equation}
where $h_0 = 7$~km.

In principle, we have to consider also contributions to the electron density from ultrarelativistic electrons and positrons produced in the (anti)muon decays. These electrons and positrons, however, appear relatively far from the anti-QN trajectory, behind the muon decay length (\ref{L-mu}). Although the total number of ions produced in the air by such electrons is comparable with the one from (anti)muons considered in this subsection, these ions are spread in a much larger volume around the anti-QN trajectory. Therefore, they give a small contribution to the electron volume density which may be neglected as compared with the ones from photons and (anti)muons.

\subsection{Total ionisation density in the initial antiquark nugget trail}

The total ionisation density of the anti-QN trail, $n_e(a,h)$, is the sum of electron volume density contributions $n_{i}(a,h)$ from all ionising particles (\ref{neSum}). Here $(n_i) = (n_\text{x},n_\gamma,n_\mu)$ denotes contributions from x-rays, $\gamma$-rays and muons, respectively. Each of these contributions is described by Eq.~(\ref{ne-i}), in which the pairs $(q_i,L_i)$ are given by Eqs.~(\ref{q-x}) and (\ref{L-x}) for x-rays, (\ref{q-gamma}) and (\ref{L-gamma}) for $\gamma$-rays and (\ref{q-mu}) and (\ref{L-mu}) for muons, respectively. 

\begin{figure}[tb]
        \centering
        \includegraphics[width=0.48\textwidth]{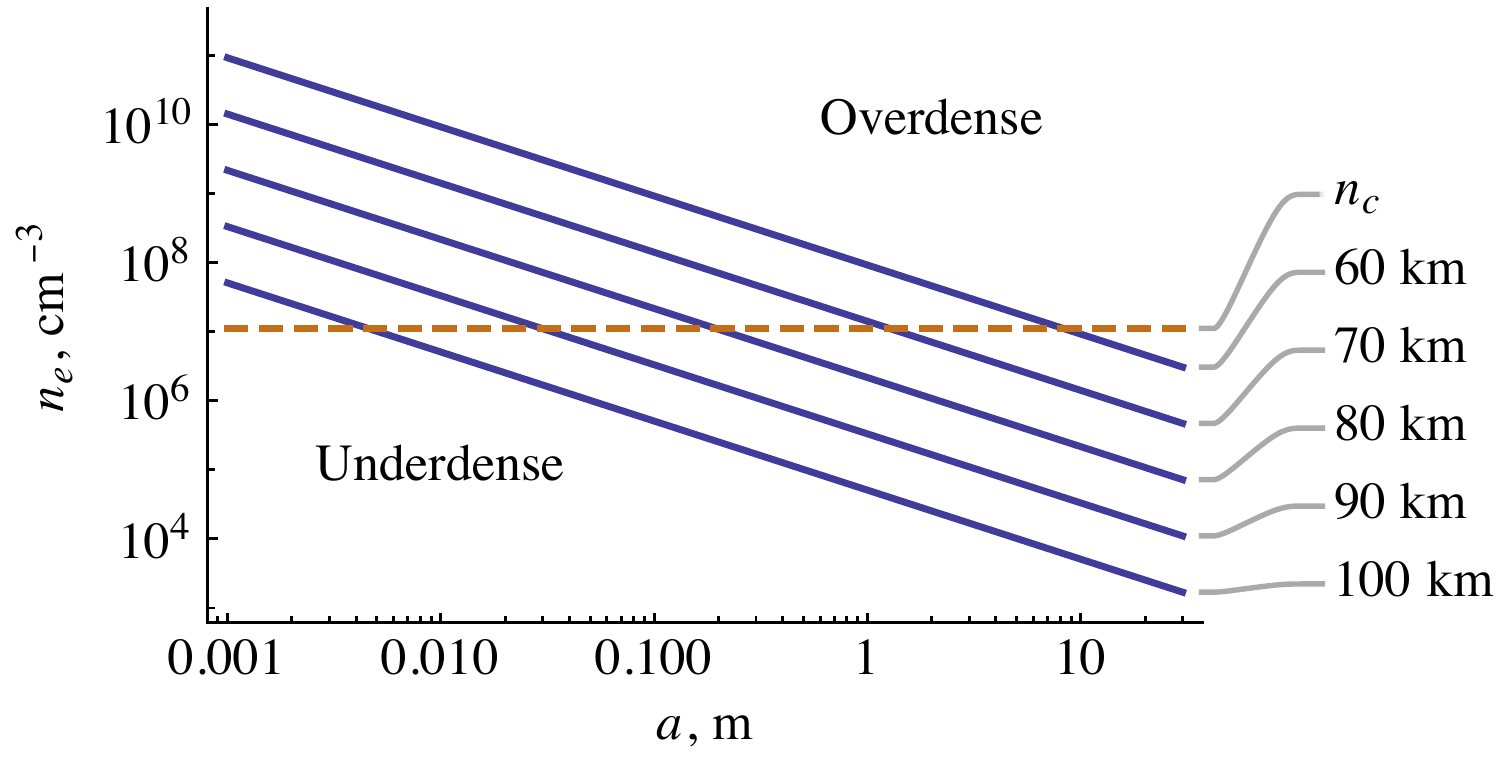}
        \caption{Electron volume density in the anti-QN trail as a function of radial distance from the anti-QN trajectory $a$. The tilted lines correspond to the electron density in the trail at different altitudes above the sea level. The horizontal dashed line is the boundary between the overdense and underdense regions, corresponding to a critical density of $ n_{c} \approx 1.1 \times 10^{7} \mbox{ cm}^{-3} $ for a typical meteor radar wavelength of $\lambda \sim 10 \mbox{ m}$.}
        \label{fig:n}
\end{figure}

It is possible to show that among different contributions to the electron density the one from x-rays dominates in the vicinity of the anti-QN trajectory at altitudes $h\gtrsim 50$ km. In this case, $n_e\approx n_\text{x}$, and for $a\ll L_\text{x}$ the electron density is approximated by the following function
\begin{equation}
    n_e(a,h) = \frac{1}{a}e^{-h/h_2}\left(\frac{7\times 10^{14}}{\text{cm}^2} \right)\,,
    \label{ne-final}
\end{equation}
where $h_2 = 5.3$~km. This function is plotted in Fig.~\ref{fig:n}. For altitudes below 50 km the general expression (\ref{neSum}) should be used.

%%%%%%%%%%%%%%%%%%%%%%%%%%%%%%%%%%%%%%%%%

\section{Detection of the anti-QN trail using meteor radars}
\label{detection}

When meteors enter the Earth's atmosphere, they produce a trail of plasma called a meteor trail. The radar detection technique of such meteor trails is well developed, see, e.g., Ref.~\cite{MeteorRadarReview} for a review. As demonstrated in the previous section, anti-QNs passing through the air also produce a trail of ionised gas. In this section, we compare the trails produced by anti-QNs with the meteor ones and show that meteor radars are suitable for detection of anti-QNs in the atmosphere. 

Note that the mechanisms of the air ionization in the meteor and anti-QN trails are different. In the former, the plasma is produced by air molecule scattering off the meteor with its subsequent meltdown and ablation. In the case of anti-QN trail, the plasma is produced by x-rays, $\gamma$-rays and fast ionizing particles appearing upon the air molecule annihilation on anti-QNs. Therefore, {\it a priori} it is not clear if the anti-QN trail may look similar to the meteor one, and whether the meteor radar detection technique could be suitable for it.

\subsection{Overdense trail width}

The working principle of meteor radar observation is the reflection of radio waves off ionised trails from meteors. The typical radar wavelength is $\lambda = 10$ m, which effectively allows one to search for meteor trails at altitudes from about 70 to 130 km above the sea level. In our estimates, we will assume the value $\lambda =10$ m for the radar wavelength, although other frequencies may be of use as well.

An important parameter in the meteor radar detection technique is the critical electron density $n_c$, which is defined as (see, e.g., \cite{MeteorRadarReview})
\begin{equation}
    n_{c} = \frac{\pi}{\lambda^{2} r_{e} } \approx 1.1\times 10^7\mbox{ cm}^{-3}\,, 
    \label{nc}
\end{equation}
where $r_e= 2.8$~fm is the classical electron radius. A region of the meteor trail with electron density below this critical density, $n_e< n_c$, is usually referred to as the underdense trail and overdense otherwise. The point about this terminology is that radiowave can penetrate inside the underdense trail, while it is fully reflected from the overdense one. As a result, the overdense parts of the trails are detected by radars with higher efficiency, while the underdense ones may be invisible if the electron density is low.

The value of the critical density (\ref{nc}) is shown in Fig.~\ref{FigTrailDensity} by the dashed line. This line allows us to find the overdense trail radius, $r_\text{overdense}$, as a solution of the equation
\begin{equation}
n_e(r_\text{overdense},h) = n_c\,,
\label{overdense-eq}
\end{equation}
where the electron density $n_e$ is given by Eq.~(\ref{neSum}) in general. 

Note that for altitudes $h \gtrsim 50$ km the electron density is given by a simple expression (\ref{ne-final}). For these altitudes, it is possible to show that the overdense trail radius obeys $r_\text{overdense}<L_\text{x}$, where $L_\text{x}$ is the x-ray absorption length, and $r_\text{overdense}$ is found analytically in this case:
\begin{equation}
    r_\text{overdense}(h) = \frac1{n_c} e^{-h/h_2} 
    \left( \frac{7\times 10^{14}}{\text{cm}^2} \right).
    \label{overdense}
\end{equation}
For lower altitudes, the overdense radius is found by solving Eq.~(\ref{overdense-eq}) numerically. The corresponding solution is plotted in Fig.~\ref{fig:trailwidth} (bottom orange line). Note that in this figure we plot the trail width which is double the trail radius, $d_\text{overdense} = 2 r_\text{overdense}$.

\subsection{Underdense trail width}

It is hard to precisely specify the lowest electron density $n_\text{min}$ in the air which may be detected because it depends on the sensitivity of a particular radar setup. The absolute low bound on the detectable electron density in the anti-QN trail is, however, natural electron density in the ionosphere, $n_\text{ion}$. This means that the electron density produced by ionizing particles and radiation from anti-QN should be at least of order of the natural electron density in the ionosphere at each particular altitude.

The natural electron density in the air $n_\text{ion}$ varies during the year and depends on the location on the Earth. In Table~\ref{tab:ionDensity} in Appendix~\ref{AppIon} we present the values of the mean electron density in the ionosphere at different altitudes from 60 to 150 km above sea level. It varies non-linearly from zero at 60 km to a few hundred thousand electrons per cm$^3$ at 150 km. 

\begin{figure}[t]
    \centering
    \includegraphics[width=0.48\textwidth]{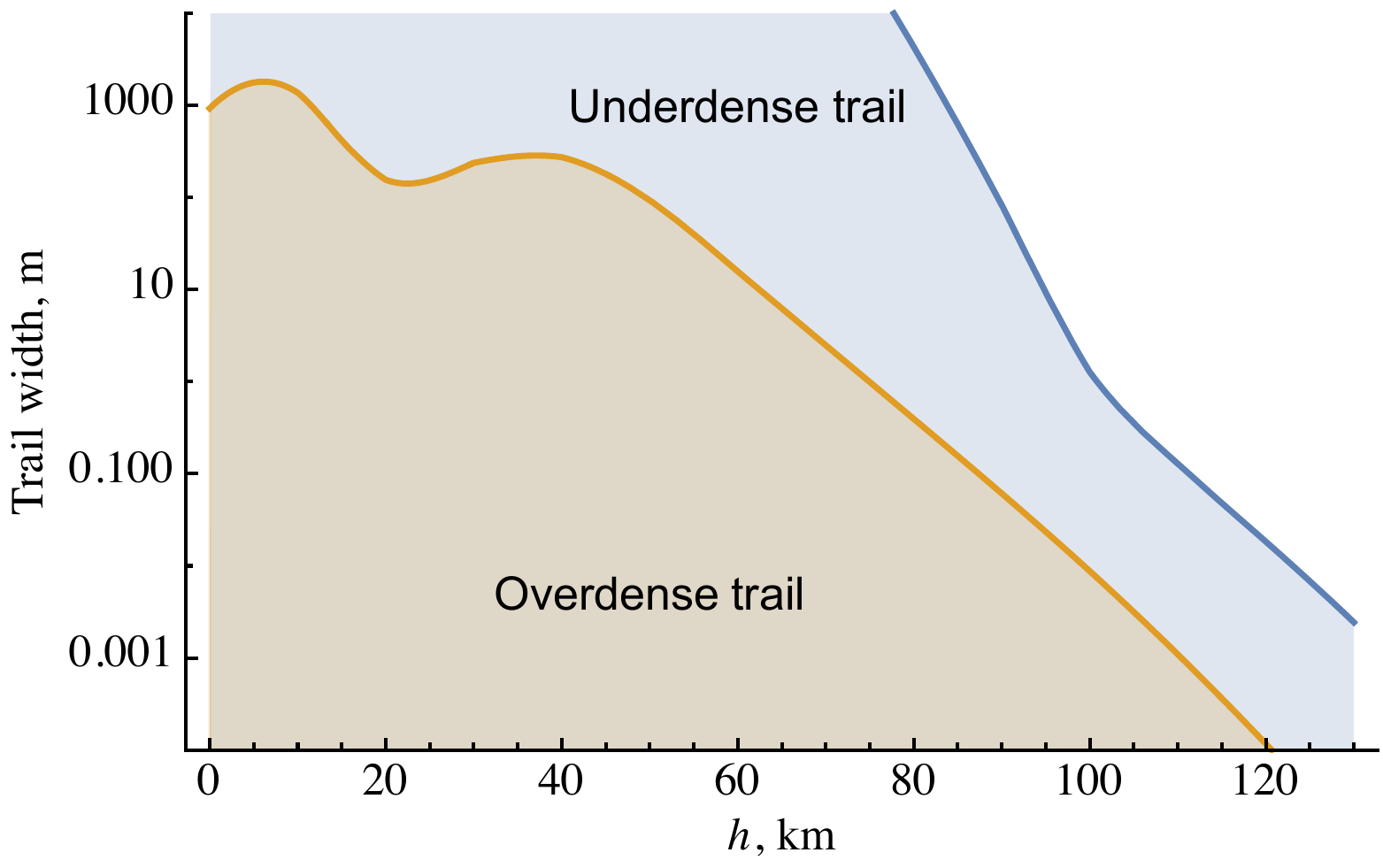}        \caption{Anti-QN trail width as a function of altitude. The overdense trail width (bottom orange line) represents a lower bound while the underdense trail width (top blue curve) gives the upper bound on the actual detectable trail width.}
        \label{fig:trailwidth}
\end{figure}

Given that the function $n_\text{ion}(h)$ is represented by numerical values in Table~\ref{tab:ionDensity}, we can solve for the equation $n_e(r_\text{underdense},h) = n_\text{ion}(h)$ to find the radius of the underdense trail at different altitudes. Making use of Eq.~(\ref{ne-final}), we find the underdense trail radius for $h\gtrsim60$ km:
\begin{equation}
r_\text{underdense}(h) = \frac1{n_\text{min}(h)}e^{-h/h_2}
\left(\frac{7\times 10^{14}}{\text{cm}^2} \right)\,.
\label{underdense}
\end{equation}
This function is represented by the top curve in Fig.~\ref{fig:trailwidth}.

Note that Eq.~(\ref{overdense}) gives a lower bound on the detectable anti-QN trail radius while Eq.~(\ref{underdense}) specifies an upper one,
\begin{equation}
    r_\text{overdense} < r < r_\text{underdense}\,.
\end{equation}
In particular at $h=100$ km, the observable trail width $d=2r$ is in the region $10\mbox{ cm}<d < 1.3\mbox{ m}$, and it grows rapidly at lower altitudes. This value is comparable with typical meteor trail width which is about 2 m at $h=100$ km \cite{JonesTrail,Kaiser2005}. Thus we conclude that the anti-QN trail may be naturally detected with standard meteor radars.

Here we considered an upper limit on the radius of underdense trail at high altitudes $h\gtrsim 60$ km, where a normal electron density is non-vanishing. For lower altitudes, it is hard to specify the underdense trail width, but it is limited by the $\gamma$-rays absorption length (\ref{L-gamma}).

Note that here we considered only the initial electron density in the anti-QN trail, which appears immediately after transition of an anti-QN through the atmosphere. We expect that the time evolution of anti-QN trails should be similar to the one in ordinary meteor trails studied in Ref.~\cite{JonesTrail}. In particular it was noted that, after a relaxation time $\tau$, the electron density in meteor trails is described by a Gaussian distribution function. This argument applies to the anti-QN trails as well. 

%%%%%%%%%%%%%%%%%%%%%%%%%%

\section{Other observable features of anti-QN trails}
\label{SecOther}

In the previous section, we demonstrated that anti-QNs produce ionized trails in the atmosphere, which are similar to the meteor ones at altitudes from about 80 to 120 km. We thus concluded that anti-QN trails may be detected with ordinary meteor radars. In this section, we discuss other features of the anti-QN trails which can help distinguish them from the meteor ones in past and future observations.

\subsection{Velocity}

The velocity of meteors entering the Earth's atmosphere ranges from 11 km/s to 72 km/s with a peak of the distribution near 30 km/s, see, e.g., \cite{MeteorVelocity}. Meteors with velocities $v>72 \mbox{ km/s}$ are classified as ones with hyperbolic trajectories in the Solar system and are believed to be of interstellar origin. Observations of such interstellar meteors are relatively rare \cite{Baggaley2000,Musci2012}.

Anti-QNs, together with QNs, are supposed to form a halo of dark matter in our Galaxy. The virial velocity of such objects near the Sun system should be on the order of 300 km/s. Thus, the expected velocity of anti-QNs in the Earth's atmosphere is much higher than the typical meteor velocity observed with radars. Unfortunately, existing interstellar meteor catalogues \cite{Baggaley2000,Musci2012} do not have records of meteors with such a high velocity, and the detection of fast interstellar meteors is challenging \cite{Hajdukova2020}. 

We have found only one reference \cite{Afanasiev2007} where a detection of an interstellar meteor with velocity about 300 km/s was reported. This meteor, however, cannot be classified as a quark nugget event, because its trail spectrum contains emissions lines of metallic elements which should be absent in the case of anti-QNs (see next subsection).

\subsection{Trail spectra}

Since meteor trails are produced from ablation, it is a well known feature of meteors that their trails possess spectral emission lines from metallic elements \cite{SpectraCatalogue}. For the anti-QN trail, however, the metallic emission lines should be absent from the trail spectra, because anti-QNs are supposed to consist of the quark rather nuclear matter. It is expected that the anti-QN trail spectrum should contain predominately emission lines of nitrogen and oxygen from ionized air molecules. 

Observations of meteor trails with missing metallic emission lines are very rare \cite{nometal}. Unfortunately, this observation cannot be identified with an anti-QN, because the observed meteor was identified as one from the Solar system with a velocity under 72 km/s. We hope that future meteor observations will reveal interstellar objects like anti-QNs with no metallic lines in their spectra.

\subsection{Frequency of antiquark nugget events}

As is noticed above, anti-QNs, if they exist, should have a very high velocity of order 300 km/s near the Earth, and their trails should be free from metallic emission lines in their spectra. Non-observation of these effects imposes limits on the frequency of anti-QN events in the Earth. The frequency of anti-QN hitting Earth  was estimated in Ref.~\cite{acoustic} assuming that QNs and anti-QNs saturate the local dark matter density in our Galaxy:
\begin{equation}
        \dot{\left< N \right>}  \simeq 2.1 \times 10^{7} \mbox{ yr}^{-1} \left( \frac{10^{25} }{\left< B \right>} \right)\,.
        \label{frequency}
\end{equation}

A typical meteor radar detection setup monitors a patch in the sky of area about $100 \times 100 \mbox{ km}^2$. Non-observation of anti-QN trails by such a setup during one year imposes a bound on the average baryon number $\langle B\rangle$ through the relation (\ref{frequency}):
\begin{equation}
    \langle B\rangle > 4\times 10^{27}\,.
    \label{Bnew}
\end{equation}
This constraint is close to the limit on the baryon number $B\gg 1.4\times 10^{27} \kappa^{3}$ obtained in Ref.~\cite{FS3} as a condition of survival of nucleons and anti-QNs in the early Universe, where parameter $\kappa$ defines the nucleon annihilation cross section relative to the geometric one, $\sigma_\text{ann}\approx \kappa \pi R_0^2$.

Note that anti-QN events may avoid detection if extremely high velocity hyperbolic meteors are excluded from observations. Hyperbolic meteors are very rare with less than 1\% of meteor velocities exceeding hyperbolic velocities \cite{mckinley1961}. Thus, anti-QN events may be overlooked in the meteor detection data.

\subsection{Upward propagating anti-QN trajectories}

Earlier limit on the baryon number of (anti)QNs was given by Eq.~(\ref{Bold}). The present study suggests a stronger limit (\ref{Bnew}) due to non-observation of specific properties of the anti-QN trail in the meteor catalogues. In our model, we assume that anti-QNs strongly interact with visible matter and annihilate air and other molecules. It is important to note that the reduction of the baryon charge of anti-QN in the proces of annihilation with air molecules is negligible. Indeed, when anti-QN crosses the Earth along its diameter it loses less than 10\% of its mass and momentum. Thus, in contrast with meteors, anti-QNs can puncture the Earth and form an up-going ionized trails in the air. Observation of such trails could be a strong evidence for the anti-QN detection in the Earth's atmosphere.

The upward propagating anti-QN trajectories are accompanied by the ionized trail, analogous to the downward moving trails considered above. We expect that these up-moving trails should be similar to ``blue jets,'' rare atmospheric events which, together with ``elves'', ``sprites'' and ``halos,'' belong to the class of transient luminous events in the upper atmosphere, see, e.g., Ref.~\cite{BlueJetsReview} for a review. These blue jets look like columns of plasma propagating from lower stratosphere up to the ionosphere. As these events are very rare, their origin is not well studied, and one could speculate about their relation to anti-QN dark matter particles. 

According to the observations, blue jets are produced only above thunderstorm clouds and represent a specific type of electrostatic discharge in the atmosphere \cite{BlueJetsReview}. The predicted anti-QN trails, however, should be unrelated to thunderstorm activity, and should be totally sporadic events on the Earth. Therefore, despite the similarity of the blue jets to the predicted upward propagating anti-QN trails, we cannot identify these two phenomena.

We conjecture that anti-QNs can initiate rare atmospheric events rather than fully explain them. If, by chance, an anti-QN passes through the region of thunderstorm activity, its ionized trail can serve as a `seed' for subsequent electric discharge in the atmosphere. A similar conjecture was advocated in Ref.~\cite{Zhitnitsky2022} to explain other exotic events in thunderstorm clouds. Such events appear when the anti-QN transitions coincides with the thunderstorm activity in a region of observations. This explains a very small frequency of observations of such events according to Ref.~\cite{Zhitnitsky2022}. 

%%%%%%%%%%%%%%%%%%%%%%%%%%%%%%%%%%%%%%%%%%%%%%%%%%%%%%%%%%%%%%%%%%%%

\section{Summary}\label{summary}

Quark nugget model of dark matter \cite{Zhitnitsky_2003} suggests that dark matter may be represented by both quark and antiquark nuggets, which consist of matter and antimatter, respectively. Anti-QNs, in contrast with QNs, strongly interact with visible matter and manifest themselves through annihilation events. In this paper, we show that annihilation products from anti-QNs create an ionized trail in the Earth's atmosphere, similar to the meteor trails. We study the properties of the anti-QN trails and compare them with meteor ones. One of the main conclusions of this paper is that the anti-QN trail could be detected using standard  meteor radars. 

Annihilation of air molecules on anti-QNs produces several types of radiation and ionizing particles: x-rays with energies in the range from 1 to 100 keV, $\gamma$ rays with energy about 200 MeV, charged pions and muons with kinetic energy about 200 MeV. We estimated the flux of these particles from anti-QNs moving through the Earth's atmosphere and studied their ionizing properties in the air. An important quantity is the electron density in the air, $n_e$, produced by all these ionizing particles and radiation from anti-QN. We found this electron density as a function of the altitude the above sea level and distance from the anti-QN trajectory. This function allows us to estimate the width of the overdense and underdense parts of the anti-QN trail, see Fig.~\ref{fig:trailwidth}. As we show, the trail width is in the range from 10 cm to 1.3 m at the 100 km altitude. This trail width is comparable with typical meteor trail width which is about 2 m at the same altitude. Thus we conclude that standard meteor radar may be suitable for detection of anti-QN trails. 

Anti-QN trails should have very specific properties which distinguish them from the meteor ones. First of all, the anti-QN velocity is on the order of 300 km/s, which is about one order in magnitude higher than the typical meteor velocity. We expect also that the spectra of anti-QN trails should be free from metallic elements emission lines, because anti-QNs do not possess atomic and nuclear structure. Meteors of small size cannot reach Earth surface while anti-QN can. Finally, if anti-QNs exist, there should be upward moving anti-QN trajectories from anti-QN particles which passed through the Earth. Currently, there have been no observations of trails which exhibit all these features. Non-observation of anti-QN trails imposes a limit on the mean baryon charge number of anti-QNs: $B>4\times 10^{27}$. DM particles with such a large baryon charge hit the Earth with the frequency less than one event per year per $100\times 100\mbox{ km}^2$ area. This constraint is close to the limit on the baryon number obtained in Ref.~\cite{FS3} as a condition of survival of nucleons and anti-QNs in the early Universe.

The above limit from non-observation of the anti-QN trails may be relaxed if some of the assumptions appear too strong. For instance, we assumed that the annihilation cross section of air molecules in collisions with anti-QNs is close to the geometric cross section $\sigma_\text{ann}\approx 0.25\pi R_0^2$. If the annihilation is strongly suppressed by some mechanism, the radiation from anti-QNs would be lower, and anti-QN trails could escape from radar meteor observations. Anyway, the limit (\ref{Bnew}) on $B$ should be confirmed by dedicated search for anti-QN trails.  It would be interesting to systematically study meteor detection catalogues with special attention to specific properties of anti-QN trails. We leave this for future works.

\vspace{3mm}
\textit{Acknowledgements} --- 
The work was supported by the Australian Research Council Grants No.\ DP230101058 and DP200100150.

%%%%%%%%%%%%%%%%%%%%%%%%%%%%%%%%%%%%%%%%%%%%%%%%%%%%%%%%%%%%%%%%%%%
\appendix
\section{Initial electron density distribution in anti-QN trail}
\label{AppA}

\begin{figure*}[tb]
    \centering
    \includegraphics[width=15cm]{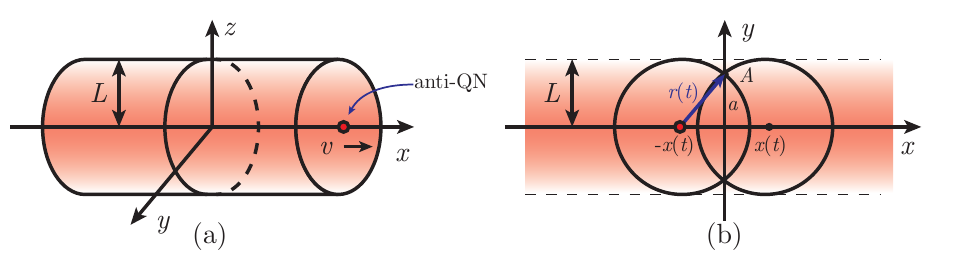}
    \caption{Electron density distribution in the anti-QN trail. Anti-QN moving with velocity $v$ in a homogeneous medium creates a cylindrical ionized trails with diameter $2L$ (left panel). Schematic contributions to the electron density in a point $A$ situated at the distance $a$ from the anti-QN trajectory (right panel).}
    \label{FigTrailDensity}
\end{figure*}

In this appendix, we estimate the electron density in the anti-QN trail $n_{e}$ as a function of distance $a$ from the anti-QN trajectory. We consider a simple model assuming that the ionized trail is created by fast relativistic particles (photons, electrons, muons, etc.) with an absorption length $L$ and ignore contributions beyond the distance $L$. We start by considering just one species of such particles assuming that different species contribute to the electron number density additively.

In general, anti-QN enters the Earth's atmosphere at a zenith angle $\theta$ and creates an ionized trail shaped as a tapered cylinder because the air density varies with the altitude. Here, however, we consider a motion of the anti-QN in a homogeneous medium (air) with a constant density $\rho_\text{air}$. It means that we neglect the variations of air density at the scale of the trail width, and the trail shape is cylindrical, see Fig.~\ref{FigTrailDensity}a. We assume also that the relaxation time in the ionized trail $\tau$ is sufficiently large, $\tau\gg v L$, where $v\sim 10^{-3}c$ is the typical anti-QN velocity. In this case, time evolution in the anti-QN trail goes relatively slowly, and we can study the initial electron density distribution $n_e$.

Let $W$ be a production rate of ionized particles in the process of air molecules annihilation on anti-QNs. Within the time interval $[t,t+dt]$, $W\, dt$ ionized particles are emitted isotropically from the anti-QN. At any given moment of time, these particles ionize the air in a spherically symmetric way, with density which drops according to the inverse square law,
\begin{equation}
    W\,dt\, n(r) = W\, dt \frac{n_0}{r^2}\,,
    \label{A1}
\end{equation}
where $n(r)=n_0/r^2$ is the electron density per one ionizing particle at distance $r$ from anti-QN, and $n_0$ is some constant. 

Let $N$ be a mean number of electrons produced by one ionizing particle along the distance $L$. This number may be written as $N={\cal E}_0/I$, where ${\cal E}_0$ is the initial (kinetic) energy of the ionizing particle, and $I$ is the mean energy required to produce one electron in the air. The latter quantity may be roughly taken as $I\approx 33$ eV for a variety of fast ionizing particles, including x-rays, $\alpha,\beta$ and $\gamma$ particles \cite{Wair}.

By integrating Eq.~(\ref{A1}) over a spherical volume $V$ of radius $L$ we find the total number of electrons in the air produced by ionizing particles during the time $dt$,
\begin{equation}
    W\,dt\, N = W\,dt\int_V n(r) d^3r = 4\pi L \,n_0 \,W\,dt\,.
\end{equation}
Thus, the constant $n_0$ may be expressed via $N$ as 
\begin{equation}
    n_0 = \frac{N}{4\pi L}\,.
    \label{n0}
\end{equation}

Consider now a point $A$ at the distance $a\leq L$ from the anti-QN trajectory (directed along the $x$-axis as in Fig.~\ref{FigTrailDensity}b). From geometric considerations it is clear that only a part of the anti-QN trajectory with $|x|\leq \sqrt{L^2 - a^2}$ should be considered in calculations of the electron density at point $A$. Then, representing $r(t) = \sqrt{x^2 + a^2} = \sqrt{v^2 t^2 + a^2}$, and integrating the expression (\ref{A1}) over the corresponding time interval we find the electron density in the point $A$:
\begin{equation}
    n_e(a) = 2\int_0^{\frac1v\sqrt{L^2-a^2}}
     \frac{W n_0\,dt}{r^2} = \frac{2Wn_0}{av}\tan^{-1}\sqrt{ \frac{L^2}{a^2} - 1}\,.
     \label{ne}
\end{equation}
With Eq.~(\ref{n0}) the electron density (\ref{ne}) becomes
\begin{equation}
    n_e(a) = \frac{WN}{2\pi L a v}\tan^{-1}\sqrt{ \frac{L^2}{a^2} - 1}\,.
    \label{ne-result}
\end{equation}
Note that $a\leq L$, and $n_e(L)=0$.

It is useful to consider also the electron line density defined as the integral of the volume density $n_e(a)$ over the ionized trail cross sectional area,
\begin{equation}
    q = 2\pi \int_0^L n_e(a) a\,da\,.
\end{equation}
This integration may be performed explicitly for the function (\ref{ne-result}):
\begin{equation}
    q = \frac{4\pi W n_0 L}{v} = \frac{WN}{v}\,.
    \label{q}
\end{equation}
As a result, the electron density (\ref{ne-result}) may be written in terms of the line density $q$ as
\begin{equation}
    n_e(a) = \frac{q}{2\pi L a }\tan^{-1}\sqrt{ \frac{L^2}{a^2} - 1}\,.
    \label{ne-q}
\end{equation}

The function (\ref{ne-q}) has non-physical behavior both at small and large values of the parameter $a$. The small-$a$ singularity of this function may be eliminated by simply taking $a\geq r_0$ where the cut-off parameter $r_0$ may be identified with anti-QN radius, $r_0\sim R_0$. At large $a$, this function should decay exponentially rather than having a sharp boundary. This may be fixed by taking the function $n(r) = n_0 e^{-r/L}/r^2$ in Eq.~(\ref{A1}). In this case, Eq.~(\ref{ne-q}) modifies as
\begin{equation}
    n_e(a) = \frac{q}{2\pi L}\int_0^\infty \frac{\exp[-\frac1L\sqrt{x^2+a^2}]}{x^2+a^2} dx\,.
    \label{ne-q+}
\end{equation}
At the intermediate values of $a$, $a<L$,
however, the functions (\ref{ne-q}) and (\ref{ne-q+}) are similar, and we prefer to use the simpler one (\ref{ne-q}) in our estimates. 

Note that in this appendix we studied the {\it initial} electron density in the anti-QN trail. Time evolution of this trail should be similar to the meteor trails studied in Ref.~\cite{JonesTrail}. In particular, it was shown that after a relaxation time $\tau$ the electron density distribution may be described by a Gaussian function regardless of the shape of the initial electron density distribution.

%%%%%%%%%%%%%%%%%%%%%%%%%%%%%%%%%%%%%%%%%%%%%%%%%%%%%

\section{Muon stopping power in air}
\label{AppMuon}

In this appendix, we collect the data on the muon stopping power and range in the air from Ref.~\cite{MuonStoppingPower}. Here we present these data for the readers' convenience.

When a muon moves through the air along the direction $x$, it loses its kinetic energy $E$ with the rate $dE/dx$, and its range may be denoted as $L$. It is convenient to normalize these quantities to the air density as follows: $\rho^{-1}_\text{air} dE/dx$ and $\rho_\text{air} L$. These normalized quantities are tabulated in Ref.~\cite{MuonStoppingPower} as functions of the muon kinetic energy $E$. In Table~\ref{TabMuon}, we present the values of these functions in the range from about 1 to 300 MeV, which is of interest in this paper. We stress that in this range the muon loses its energy predominantly to the air molecule ionization. Therefore, these data are suitable for calculations of the electron density produced by near-relativistic muons moving through the air.

\begin{table}[tbh]
    \centering
    \begin{tabular}{c|c|c}
    $E$, MeV & $\rho^{-1}_\text{air} dE/dx$, MeV\,cm$^2$/g & $\rho_\text{air} L$, g/cm$^2$ \\\hline
  1.2 & 38.30 & $7.018\times 10^{-3}$  \\
%  1.7 & 28.89 & $2.222\times 10^{-2}$  \\
  2.0 & 25.32 & $3.334\times 10^{-2}$  \\
%  2.5 & 21.12 & $5.506\times 10^{-2}$  \\
  3.0 & 18.22 & $8.063\times 10^{-2}$  \\
%  3.5 & 16.08 & $1.099\times 10^{-1}$  \\
  4.0 & 14.44 & $1.428\times 10^{-1}$  \\
%  4.5 & 13.14 & $1.791\times 10^{-1}$  \\
  5.0 & 12.08 & $2.188\times 10^{-1}$  \\
%  5.5 & 11.19 & $2.619\times 10^{-1}$  \\
  6.0 & 10.45 & $3.081\times 10^{-1}$  \\
%  7.0 & 9.258 & $4.100\times 10^{-1}$  \\
  8.0 & 8.346 & $5.240\times 10^{-1}$  \\
%  9.0 & 7.625 & $6.495\times 10^{-1}$  \\
  10 & 7.039 & $7.862\times 10^{-1}$  \\
%  12 & 6.145 & 1.091  \\
  14 & 5.495 & 1.436  \\
  17 & 4.793 & 2.023  \\
  20 & 4.294 & 2.686  \\
  25 & 3.720 & 3.942  \\
  30 & 3.333 & 5.366  \\
%  35 & 3.056 & 6.936  \\
  40 & 2.847 & 8.633  \\
%  45 & 2.686 & 10.44  \\
  50 & 2.558 & 12.35  \\
%  55 & 2.454 & 14.35  \\
  60 & 2.368 & 16.42  \\
%  70 & 2.236 & 20.78  \\
  80 & 2.140 & 25.35  \\
%  90 & 2.068 & 30.11  \\
  100 & 2.014 & 35.01  \\
%  120 & 1.937 & 45.15  \\
  140 & 1.889 & 55.62  \\
%  170 & 1.848 & 71.69  \\
  200 & 1.827 & 88.03  \\
  250 & 1.816 & 115.5  \\
  300 & 1.819 & 143.0  \\\hline
    \end{tabular}
    \caption{Muon stopping power and range in the dry air at 1 atm with density $\rho_\text{air} = 1.205\times 10^{-3}$ g/cm$^3$. Data are taken from Ref.~\cite{MuonStoppingPower}.}
    \label{TabMuon}
\end{table}

\section{Average electron density in the ionosphere}
\label{AppIon}

Normal density of electrons in the ionosphere varies daily and annually, as well as it depends on the location on the Earth. In our work, it is sufficient to consider the mean electron density in the ionosphere presented, e.g., Ref.~\cite{IonosphereBook}. This density is a non-linear function of altitude.  Numerical values of this function at the altitudes of interest are given in Table~\ref{tab:ionDensity}.

\begin{table}[tbh]
    \centering
    \begin{tabular}{c|c}
    $h$, km & $n_\text{ion}$, cm$^{-3}$\\\hline
 60 & 80 \\
 70 & 200 \\
 80 & 1000 \\
 90 & 8000 \\
 100 & $8\times 10^{4}$ \\
 110 & $12\times 10^{4}$ \\
 120 & $13\times 10^{4}$ \\
 130 & $15\times 10^{4}$ \\
 140 & $22\times 10^{4}$ \\
 150 & $30\times 10^{4}$ \\\hline
    \end{tabular}
    \caption{Mean electron number density in the ionosphere. Data taken from  Ref.~\cite{IonosphereBook}.}
    \label{tab:ionDensity}
\end{table}

%%%%%%%%%%%%%%%%%%%%%%%%%%%%%%%%%%%%%%%%%%%%%%%%%%%%%%%%%%%%%%%%%%%

%apsrev4-2.bst 2019-01-14 (MD) hand-edited version of apsrev4-1.bst
%Control: key (0)
%Control: author (72) initials jnrlst
%Control: editor formatted (1) identically to author
%Control: production of article title (-1) disabled
%Control: page (0) single
%Control: year (1) truncated
%Control: production of eprint (0) enabled
%

\end{document}